# Panorama do mercado global da indústria de semicondutores

# Overview of the global semiconductor industry market

**Clóves Gonçalves Rodrigues**
Doutor em Ciências pelo Instituto de Física Gleb Wataghin, Unicamp, SP, Brasil
Instituição: Pontifícia Universidade Católica de Goiás – PUC Goiás
Endereço: Av. Universitária, n. 1440, CP 86, Setor Universitário, Goiânia-GO, 74605-010
E-mail: cloves@pucgoias.edu.br

**RESUMO**
Dominar a tecnologia de semicondutores é fundamental para inserir qualquer país nas tendências do futuro, tais como: cidades inteligentes, internet das coisas, exploração espacial, etc. Neste trabalho comentamos sobre a importância do domínio sobre a tecnologia de produção de semicondutores e as suas implicações no desenvolvimento de uma nação e apresentamos o crescente faturamento global da indústria de semicondutores nos últimos 20 anos.

**Palavras-Chave:** Indústria de Semicondutores, Mercado Global, Internet das Coisas, Cidades Inteligentes.

**ABSTRACT**
Mastering semiconductor technology is essential to insert any country into the trends of the future, such as smart cities, internet of things, space exploration, etc. In this paper we present the growing annual revenue of the semiconductor industry in the last 20 years and comment on the importance of mastering semiconductor production technology and its implications for the development of a nation.

## 1 INTRODUÇÃO

Para melhorar a sua condição de vida, o ser humano sempre procurou transformar os recursos naturais disponíveis em produtos cada vez mais eficientes, criando novas possibilidades de aplicações. Dessa forma, os materiais passaram a ser classificados conforme as suas principais características, o que possibilitou classificar as propriedades interessantes e comuns dos materiais para as mais variadas aplicações (Braithwaite, 1990), (Swart, 2008), (Dias, 2005).

Materiais são substâncias cujas propriedades os tornam utilizáveis em máquinas, dispositivos, estruturas, ou produtos consumíveis. Assim, é fundamental conhecer as propriedades dos materiais para uma correta aplicação dos mesmos conforme as suas características. Exemplos de materiais com propriedades distintas são: cerâmicas, metais, supercondutores, semicondutores, plásticos, polímeros, fibras, vidros, madeira, areia, pedra, conjugados, etc. Em quase todas as atividades humanas existe uma certa dependência em relação à utilização de materiais. Esses materiais são utilizados nas residências, nos meios de transporte, em meios de comunicação, no vestuário, no comércio, no lazer, no processamento de dados, na produção de alimentos, no ensino, na saúde, na produção e distribuição de energia, e em muitas outras áreas, atividades e segmentos. Assim, o conhecimento e a capacidade em produzir e manejar materiais afetam de forma direta a qualidade de vida da população (Braithwaite, 1990), (Swart, 2008), (Dias, 2005). Por isto, dominar o conhecimento dos princípios básicos e fundamentais dos materiais é também de extrema importância para que o desenvolvimento de novos materiais e processos de fabricação possam ter um progresso contínuo.

Historicamente nota-se que o nível de desenvolvimento de um povo está estreitamente ligado à sua habilidade em fabricar e manipular os materiais. Dessa forma, as culturas passadas foram classificadas de acordo com tal habilidade como, por exemplo, a "idade da pedra", a "idade do bronze" a "idade do ferro", etc. Em analogia, estamos atualmente no que poderíamos chamar de "idade da eletrônica" (Braithwaite, 1990), (Swart, 2008), (Dias, 2005).

Atualmente a maioria das atividades humanas dependem, de forma direta ou indireta, de algum sistema eletrônico. Os componentes eletrônicos estão presentes na grande maioria das atividades de uma sociedade como, por exemplo, na comunicação, no transporte, na computação, no controle de processos industriais, em instrumentos de pesquisa e análise, e em inúmeras outras atividades. Assim, a eletrônica está se tornando o maior mercado mundial, tendo um valor atual estimado acima de um trilhão de dólares, lembrando que todos os dispositivos eletrônicos são baseados em algum tipo de material (Braithwaite, 1990), (Swart, 2008), (Dias, 2005).

Sendo a propriedade elétrica uma característica muito importante dos materiais é de extrema importância a pesquisa e o estudo desta propriedade. A condutividade elétrica é o caráter elétrico de um material, ou seja, sua capacidade de conduzir corrente elétrica. Do ponto de vista da propriedade da condutividade elétrica os materiais podem ser agrupados em: isolantes, condutores, supercondutores e semicondutores. Entre os diversos grupos de materiais com propriedades elétricas interessantes destacam-se os semicondutores (Pereira, 2021; RODRIGUES, 2006, 2007, 2008, 2009, 2012, 2017, 2018, 2021). Os materiais semicondutores provocaram um formidável impulso na indústria eletrônica e promoveram uma grande revolução da computação e da eletrônica (Rodrigues, 2010). Dentre os dispositivos criados com materiais semicondutores podemos citar: diodos, diodos emissores de luz (LEDs), transistores, detectores e emissores diversos, placas solares, sensores, lasers, etc.

## 2 O MERCADO MUNDIAL DA INDÚSTRIA DE SEMICONDUTORES

A tecnologia de semicondutores pode ser considerada há pelo menos 150-200 anos atrás, quando cientistas e engenheiros que moravam às margens ocidentais do Lago Victoria produziam aço-carbono feito de cristais de ferro em vez de sinterização de partículas sólidas (Henisch, 1968). Desde então, o silício e materiais à base de silício têm desempenhado um papel importante no desenvolvimento da moderna tecnologia de dispositivos semicondutores. Outros materiais semicondutores foram descobertos ao longo do tempo sendo continuamente incorporados à indústria de semicondutores.

A indústria de semicondutores é uma das mais relevantes e dinâmicas do mundo atual. A sua importância está relacionada às crescentes possibilidades de aplicações dos componentes semicondutores, que não se restringem mais às indústrias eletrônicas e de telecomunicações. Os semicondutores estão no âmago não apenas da "Era da Informação e da Comunicação", como também da Indústria 4.0, também conhecida como a "Revolução da Internet das Coisas". A importância da indústria de semicondutores também está relacionada ao significativo impacto que a evolução tecnológica do setor tem sobre a produtividade de todos os demais setores da economia, uma vez que a crescente capacidade de processamento e armazenamento foi acompanhada por um custo decrescente.

A Figura 1 mostra o faturamento anual da indústria mundial de semicondutores desde o ano de 2001 (Gartner, 2021). Nota-se que este faturamento cresceu ano após ano (com excessão dos anos de 2009 e 2019). Tomando, como exemplo, o ano de 2018 este faturamento foi de 476,7 bilhões de dólares, enquanto que de acordo com o IBGE o Produto Interno Bruto Brasileiro (PIB) em 2018 foi de 6,8 trilhões de dólares (Advfn, 2020). Assim, o faturamento mundial da indústria de semicondutores representa cerca de 7% em relação ao PIB nacional de 2018.

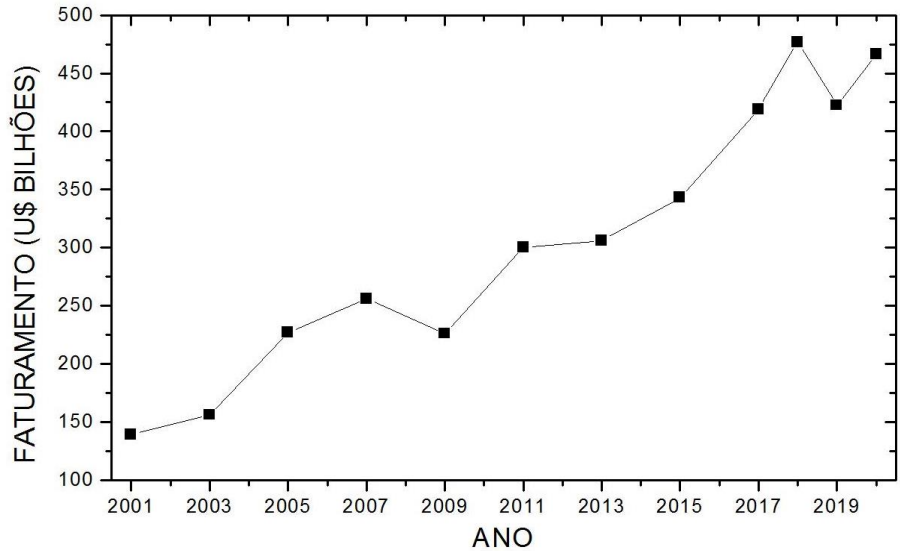

Figura 1. Faturamento anual em bilhões de dólares pela indústria de semicondutores.
Fonte: (Gartner, 2021).

No ano de 2019 o mercado de semicondutores foi impactado por vários fatores. Um ambiente de preços fracos para memórias e outros tipos de chips, o baixo crescimento das principais aplicações dos semicondutores como smartphones, servidores e PCs, e uma disputa comercial entre os Estados Unidos e a China conduzindo o mercado para o mais baixo crescimento desde 2009 (Reuters, 2020). Existem diversas restrições que os Estados Unidos impõem aos negócios chineses, as quais são baseadas em preocupações de segurança nacional. Segundo especialistas em finanças e economia, os gerentes de produtos semicondutores deveriam revisar as estratégias de produção e os planos de investimentos para se proteger desses eventuais períodos em que os mercados se tornam mais fracos (Ipesi, 2020).

Em 2020 a receita mundial de semicondutores totalizou US$ 466,2 bilhões. De acordo com os resultados apurados pelo Gartner, houve um crescimento de 10,4% em relação ao ano de 2019. A venda de semicondutores (ou processadores) é um termômetro no mercado de equipamentos, pois sinaliza a atividade de produção.

Segundo Andrew Norwood, vice-presidente de pesquisa do Gartner, o crescimento da indústria de semicondutores foi liderada por Memórias, GPUs (Graphics Processing Unit, ou Unidade de Processamento Gráfico) e Chipsets 5G, impulsionados pela demanda do mercado final de celulares hyperscale, PCs, ultramóveis e 5G. No entanto, a eletrônica industrial e automotiva sofreram em 2020 uma pausa ou redução no consumo de semicondutores devido à pandemia provocada pela COVID-19.

O primeiro lugar no ranking em receita no ano de 2020 como fornecedor global de semicondutores foi a Intel, sendo seguida por: Samsung Electronics, SK hynix e Micron (veja a Tabela 1). A receita de semicondutores da Intel cresceu 7,4% em relação a 2019, impulsionada pelo crescimento de seus principais negócios de CPU, tanto de clientes quanto de servidores. No geral, os melhores desempenhos entre os 10 primeiros da lista foram da NVIDIA e da MediaTek. O crescimento de 45,2% da NVIDIA foi impulsionado principalmente pelos negócios relacionados a jogos da empresa e em data centers. Aproveitando o vácuo deixado pela concorrente direta Huawei, que teve os seus negócios interrompidos ao longo do ano de 2020, a receita da MediaTek cresceu 38,1% em 2020 *(Gartner, 2021)*.

Tabela 1. Top 10 Global em 2020 dos principais fornecedores de semicondutores por receita (em milhões de dólares americanos).

| Posição (2019) | Posição (2020) | Empresa Fornecedora | Vendas (2019) | Vendas (2020) | Cota de Mercado em 2020 | Crescimento (2019-2020) |
|---|---|---|---|---|---|---|
| 1º | 1º | Intel | 67.754 | 72.759 | 15,6% | 7,4% |
| 2º | 2º | Samsung Electronics | 52.389 | 57.729 | 12,4% | 10,2% |
| 3º | 3º | SK hynix | 22.297 | 25.854 | 5,5% | 16,0% |
| 4º | 4º | Micron Technology | 20.254 | 22.037 | 4,7% | 8,8% |
| 6º | 5º | Qualcomm | 13.613 | 17.632 | 3,8% | 29,5% |
| 5º | 6º | Broadcom | 15.322 | 15.754 | 3,4% | 2,8% |
| 7º | 7º | Texas Instruments | 13.364 | 13.619 | 2,9% | 1,9% |
| 13º | 8º | MediaTek | 7.958 | 10.988 | 2,4% | 38,1% |
| 16º | 9º | NVIDIA | 7.331 | 10.643 | 2,3% | 45,2% |
| 14º | 10º | KIOXIA | 7.827 | 10.374 | 2,2% | 32,5% |
|  |  | Outras (fora do top 10) | 194.228 | 208.848 | 44,8% | 7,5% |
|  |  | Total | 422.337 | 466.237 | 100% | 10,4 |

Fonte: Gartner (Abril 2021).

A segunda categoria de dispositivos com uma melhor receita, com um aumento de 13,5% em relação ao ano de 2019 foi a memória, respondendo por 26,7% das vendas de semicondutores em 2020. O crescimento na produção de memórias se deve a principal tendência do ano de 2020: a mudança para aprendizado (aulas online) e trabalho em casa (home office) além do entretenimento online. Isto alimentou o aumento da construção de servidores de fornecedores de hiperescala para satisfazer este comportamento, bem como um aumento pela demanda de PCs e ultramobiles. Ainda em relação à memória, o flash NAND obteve o melhor desempenho com um crescimento de receita de 25,2% devido a uma escassez no ocorrida no primeiro semestre de 2020. A tendência é que no ano de 2021, tanto flash NAND quanto DRAM estarão em falta, aumentando os preços ao longo do ano e as receitas disparando em aproximadamente 25%. Estima-se que a receita da indústria de semicondutores vai continuar tendo um crescimento composto anual nos próximos três anos. (IDC, 2021).

## 3 COMENTÁRIOS FINAIS

Os semicondutores são a base para a fabricação dos processadores que controlam computadores, smartphones, eletrodomésticos, e praticamente todo e qualquer aparelho moderno. Trata-se de um produto que requer tecnologia de ponta para ser fabricado. Portanto, dominar a tecnologia e a fabricação de materiais semicondutores é muito importante. Com a eletrônica cada vez mais presente nos objetos e produtos que nos acompanham diariamente e com o advindo das tão esperadas cidades inteligentes e o crescimento da internet das coisas, não é interessante para nenhum país viver na dependência de produtos semicondutores importados.

A tecnologia de semicondutores é uma tecnologia muito complexa. Ela envolve diversas áreas de engenharias além de uma eletrônica com processos robotizados e automatizados, com muito projeto e tecnologia embarcados necessitando de um alto conhecimento de funcionários e recursos humanos.

A maioria das máquinas em uma indústria de semicondutores são operadas a distância e estão conectadas entre si. Todo o processo é acompanhado em tempo real, desde a qualidade do produto, a quantidade da produção, e até mesmo a especificação de cada etapa, exigindo um grande nível de tecnologia, automação e controle.

Atualmente, a grande maioria dos componentes de equipamentos eletrônicos, de baterias a processadores, são produzidos em países localizados na Ásia. Isto acontece porque além da mão de obra qualificada, o custo de produção no continente asiático é mais baixo. É muito difícil pra qualquer país concorrer neste segmento com os asiáticos.

Um dos problemas que surge com isto é que a utilização de componentes semicondutores importados da Ásia pela indústria de eletrônicos deixa tudo mais caro para o consumidor final e o país continua dependente da tecnologia de outras nações. O Brasil, por exemplo, é um país rico em minérios. Porém, no Brasil não existem empresas especializadas no seu tratamento, como o seu crescimento e os processos de preparação de lâminas para a indústria de semicondutores. O resultado disto é que o Brasil gasta anualmente cerca de 6 bilhões de dólares com a importação de semicondutores. O domínio sobre a tecnologia de produção de semicondutores é de vital importância para qualquer país do mundo para que o mesmo não tenha que ser inteiramente dependente dos asiáticos.

Dominar a tecnologia de semicondutores é fundamental para inserir qualquer país nas tendências do futuro, como cidades inteligentes, internet das coisas, exploração espacial, etc. Finalizando, fechamos com as palavras do físico Marcelo Gleiser: "Ciência não é uma escolha. É uma necessidade. Quem não encherga issso está fadado a um obscurantismo que condena o futuro do país" (GLEISER, 2021).

# REFERÊNCIAS